\documentclass[preprint]{iucr}              

     \journalcode{}              

\usepackage{amsmath}
\usepackage{amsfonts}
\usepackage{graphicx}
\usepackage{color}

\begin{document}                  



\title{Random Conical Tilt Reconstruction without Particle Picking in Cryo-electron Microscopy}


\cauthor[a]{Ti-Yen}{Lan}{tiyenlan@princeton.edu}{}
\author[b]{Nicolas}{Boumal}
\author[a,c]{Amit}{Singer}

\aff[a]{Program in Applied and Computational Mathematics, Princeton University, Princeton, NJ 08544, \country{USA}}
\aff[b]{Institute of Mathematics, EPFL, CH-1015 Lausanne, \country{Switzerland}}
\aff[c]{Department of Mathematics, Princeton University, Princeton, NJ 08544, \country{USA}}









\maketitle                        

\begin{synopsis}
We describe a method to reconstruct the 3-D molecular structure without the need for particle picking in the random conical tilt scheme in cryo-electron microscopy. Our results show promise to reduce the size limit for single particle reconstruction in cryo-electron microscopy.
\end{synopsis}

\begin{abstract}
We propose a method to reconstruct the 3-D molecular structure from micrographs collected at just one sample tilt angle in the random conical tilt scheme in cryo-electron microscopy. Our method uses autocorrelation analysis on the micrographs to estimate features of the molecule which are invariant under certain nuisance parameters such as the positions of molecular projections in the micrographs. This enables us to reconstruct the molecular structure directly from micrographs, completely circumventing the need for particle picking. We demonstrate reconstructions with simulated data and investigate the effect of the missing-cone region. These results show promise to reduce the size limit for single particle reconstruction in cryo-electron microscopy.
\end{abstract}

\section{Introduction}
\begin{figure}
\label{fig:conventional-RCT}
\includegraphics[scale=0.27, trim=0cm 0cm 0cm 0cm, clip=true]{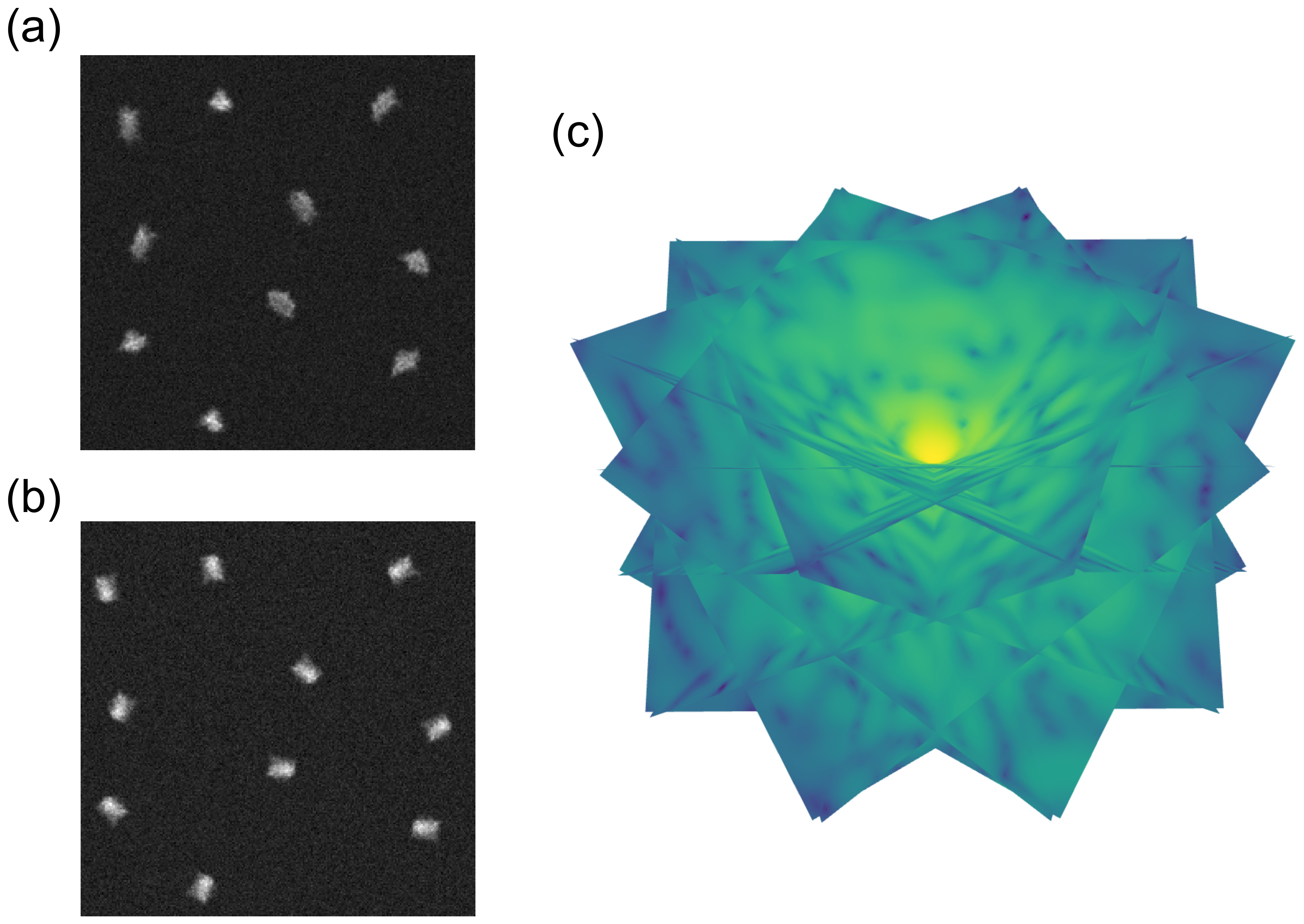}
\caption{The micrographs of the same field of view collected at (a) one large sample tilt angle and (b) no tilt. (c) The Fourier transforms of the molecular projections recorded in (a), which are assembled in Fourier space with respect to their corresponding orientations according to the Fourier slice theorem discussed in Section~\ref{sec:rct}.}
\end{figure}

Random conical tilt (RCT)~\cite{Radermacher1987,Radermacher1988} is an important technique in single-particle cryo-electron microscopy (cryo-EM) to generate a \textit{de novo} 3-D reconstruction, which provides an unbiased initial model for a subsequent iterative refinement process to determine high-resolution structures. The technique applies to molecules that have a preferred orientation to the 2-D substrate they are deposited on and random in-plane rotations. The standard data collection scheme of RCT involves measuring pairs of images, or micrographs, of the same field of view: one with a large sample tilt angle (Figure~\ref{fig:conventional-RCT}(a)), and one with no tilt (Figure~\ref{fig:conventional-RCT}(b)). Since the micrograph pairs contain projections of each molecule at two views that are physically related, one can first estimate the in-plane rotation of each molecule by aligning the molecular projections measured in the untilted micrographs and then assemble the corresponding molecular projections recorded in the tilted micrographs to reconstruct the 3-D molecular structure, as shown in Figure~\ref{fig:conventional-RCT}(c). 

However, some limitations exist for the RCT method. The design of the sample holder restricts the maximum tilt angle to about $60^\circ$, which makes a considerable fraction of information about the molecular structure inaccessible to the technique: this is the so-called ``missing-cone'' problem. Another limitation is the need to collect data from the same field of view at two different sample tilt angles. For each of the two tilt angles, the signal-to-noise ratio (SNR) must be high enough so that it is possible to reliably locate the molecular projections (that is, pick particles) in the noisy micrographs. This essentially doubles the required electron dose on the sample. Meanwhile, the molecule must be large enough so that the irreversible structural damage caused by incident electrons is limited enough to allow for particle picking. Indeed, this has led to the common belief that small biological molecules are out of the reach for cryo-EM~\cite{Henderson1995}.

\begin{figure}
\label{fig:schematics}
\includegraphics[scale=0.27, trim=0cm 0cm 0cm 0cm, clip=true]{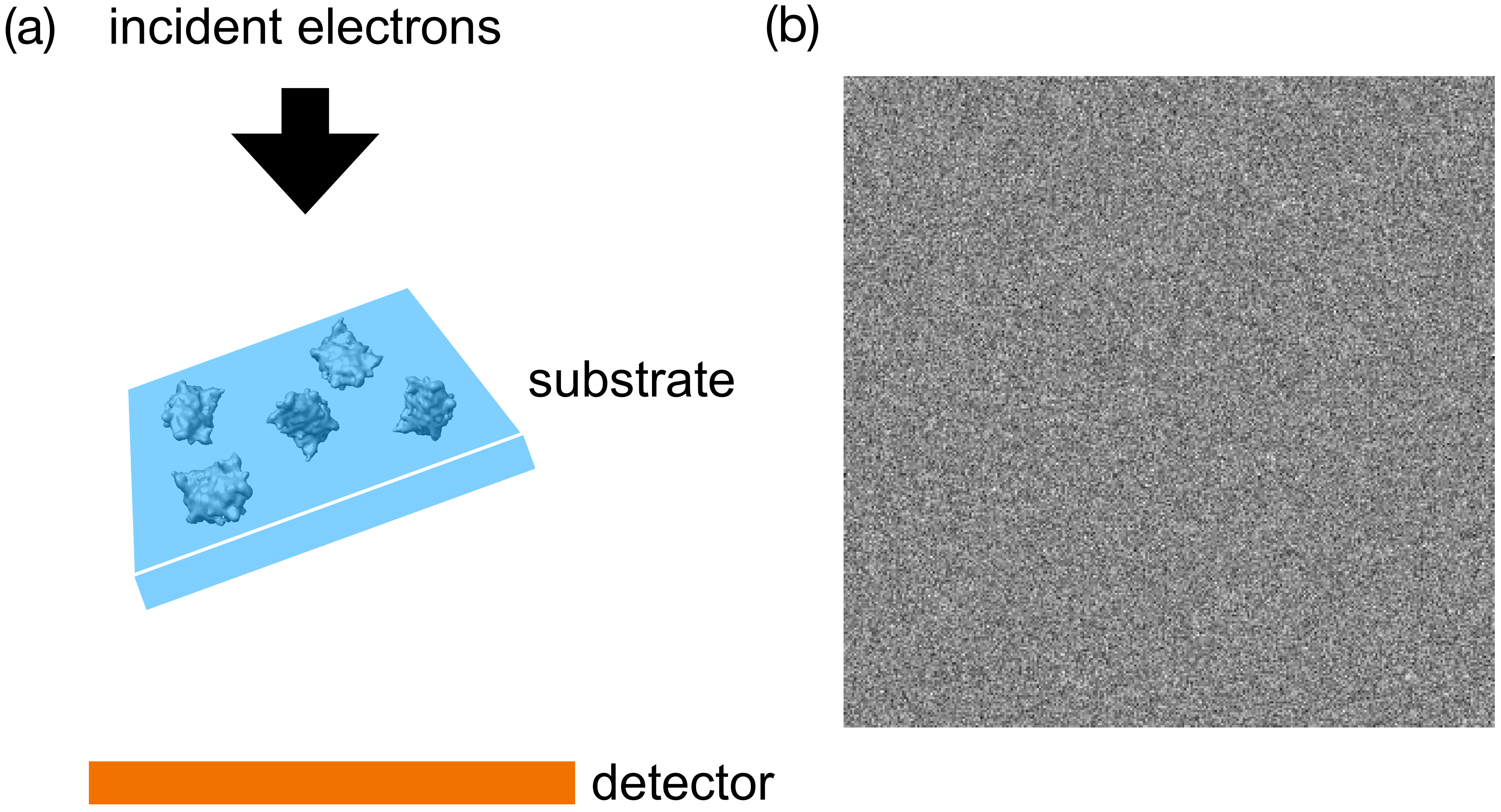}
\caption{(a) The data collection scheme of RCT with just one sample tilt angle. (b) A micrograph that is so noisy that picking particles is challenging.}
\end{figure}

In this study, we develop an approach to reconstruct the 3-D molecular structure from data collected at just one large sample tilt angle, as depicted in Figure~\ref{fig:schematics}(a). More importantly, our approach circumvents the need for particle picking to reconstruct the molecular structure directly from the micrographs. The main idea is to first estimate features of the molecule that are invariant to the 2-D positions of molecular projections in the micrographs. The estimation is done through a variant of Kam's autocorrelation analysis~\cite{Kam1980}. We subsequently determine the molecular structure by fitting the estimated invariants through an optimization problem. We address the problem of missing information by adding a regularizer in the optimization. Assuming white noise, this approach can in principle handle cases of arbitrarily low SNR as long as sufficiently many micrographs are used to estimate the invariants. Figure~\ref{fig:schematics}(b) shows one such noisy micrograph where particle picking becomes challenging. This observation notably suggests that the feasibility of particle picking does not limit the smallest usable molecule size in single-particle cryo-EM. 

Kam's autocorrelation analysis was also applied for analyzing X-ray single particle imaging data~\cite{Kam1977,Saldin2010,Donatelli2015,vonArdenne2018}. In particular, \citeasnoun{Saldin2010} considered the problem of reconstructing the top-down projection of molecules randomly oriented about a single axis, which is similar to the case of no tilt in RCT. Subsequently, \citeasnoun{Elser2011} designed an algorithm to reconstruct the 3-D structure of such partially oriented molecules from a tilt series. Kam's method was recently demonstrated with actual data collected from randomly oriented virus particles~\cite{Kurta2017}. 

This work belongs to a methodical program to develop algorithms to reconstruct molecular structures without the need for particle picking, which was first proposed in \citeasnoun{Bendory2018}. The development started with the studies of a simplified 1-D model, where multiple copies of a target signal occur at unknown locations in a noisy long measurement~\cite{Bendory2018,Bendory2019,Lan2020}. The extension to the 2-D case, where multiple copies of a target image are randomly rotated and translated in a large noisy measurement image, was later studied in \citeasnoun{Marshall2020} and \citeasnoun{Bendory2021}. These results can be used to reconstruct the top-down molecular projection from the micrographs collected at no tilt in the RCT scheme. 

We organize the rest of the paper as follows. We describe the data simulation procedure in Sections~\ref{sec:img_form} to \ref{sec:micrograph}. The details of our approach are discussed in Sections \ref{sec:autocorr} and \ref{sec:opt}. In Section \ref{sec:results}, we study the effect of the missing-cone region on the quality of reconstruction and present the reconstructions of two molecular structures from simulated noisy micrographs. The computational details are described in the appendix.

\section{Methods}
\subsection{Image formation model}
\label{sec:img_form}
In the cryo-EM imaging process, the incident electrons are scattered by the \mbox{3-D} Coulomb potential of the sample $f_s(x, y, z)$. We define the coordinate system for data collection $S$ by the orthogonal $\mathbf{x}$- and $\mathbf{y}$-axes along the edges of the detector and the normally incident electron beam, as the $\mathbf{z}$-axis. Under the weak-phase object approximation, the micrograph recorded by an $m\times m$ pixelated detector can be modeled as
\begin{equation}
M(x_i, y_i) = \left( h \ast \mathcal{P} f_s \right) (x_i/\xi, y_i/\xi) + \varepsilon(x_i, y_i), 
\end{equation}
where $i \in \{1, \dots, m^2\}$, $(x_i, y_i) \in \{ -\lfloor m/2 \rfloor, \dots, \lceil m/2 - 1 \rceil \}^2$ is the 2-D coordinate of the $i^\mathrm{th}$ pixel, and $\xi$ denotes the pixel sampling rate. The operator $\mathcal{P}$ generates the tomographic projection of $f_s$ along the $\mathbf{z}$-axis by
\begin{equation}
(\mathcal{P}f_s) (x, y) = \int_{-\infty}^{\infty} f_s(x, y, z)~dz.
\end{equation}
The 2-D function $h(x, y)$ represents the point spread function of the imaging system, and the operator $\ast$ denotes the 2-D convolution, where
\begin{equation}
(h\ast g) (x, y) = \int_{-\infty}^{\infty} \int_{-\infty}^{\infty} h(u, v) g(x-u, y-v)~du~dv
\end{equation}
for any 2-D function $g(x, y)$. Finally, the measurement noise is modeled by the additive random variable $\varepsilon(x_i, y_i)$. 

In this work, we consider the simplified scenario where we ignore the effect of the point spread function by making the idealistic assumption that it is a 2-D Dirac delta function, namely, $(h\ast g) (x, y) = g(x,y)$. Moreover, we assume that the random noise $\varepsilon$ is drawn from an i.i.d. Gaussian distribution with zero mean and variance $\sigma^2$. The arising challenges beyond these assumptions will be discussed in~Section~\ref{sec:discussion}.

\subsection{Random conical tilt}
\label{sec:rct}
\begin{figure}
\label{fig:coordinates}
\includegraphics[scale=0.27, trim=15cm 6cm 10cm 5cm, clip=true]{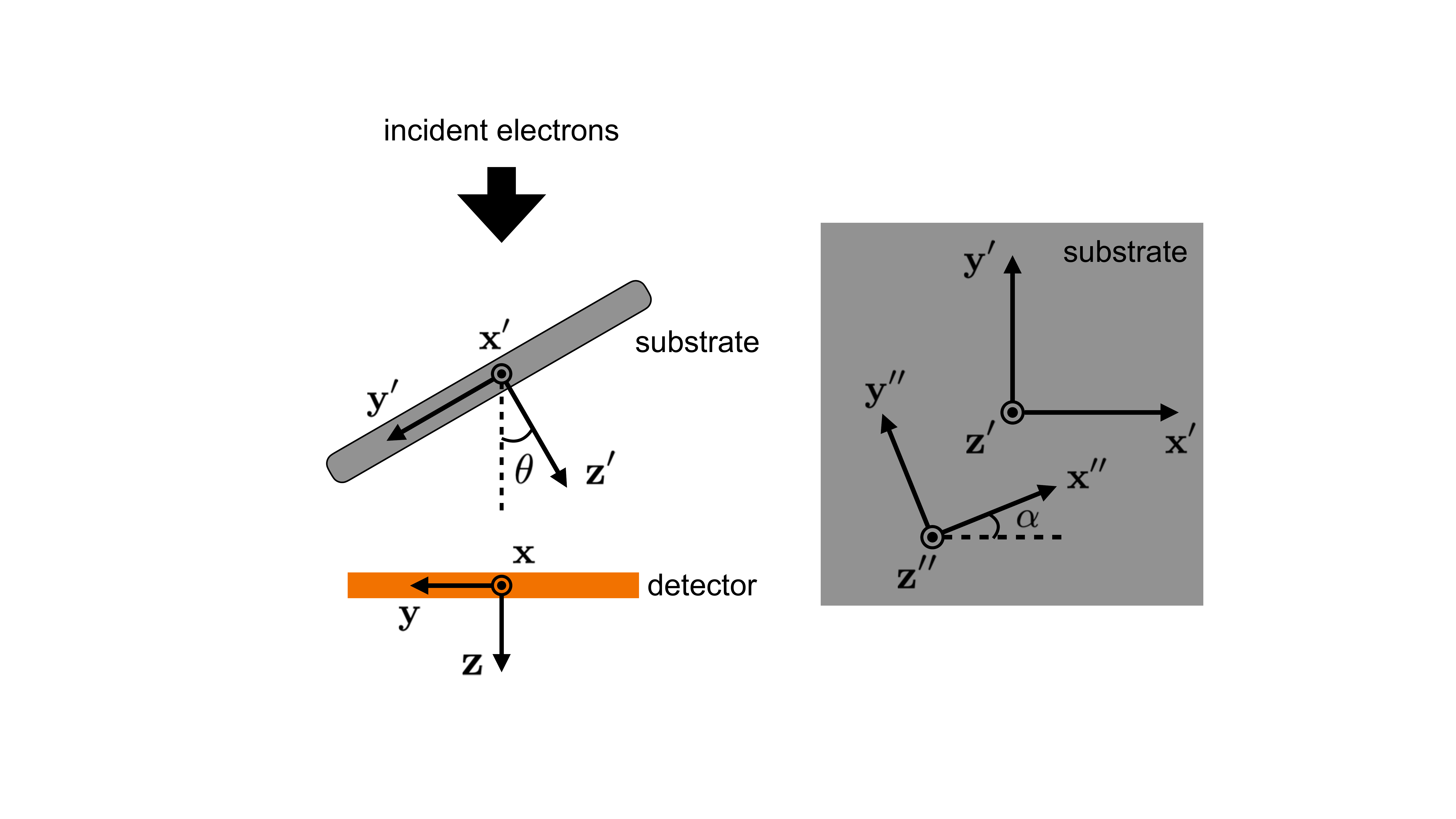}
\caption{The relationships between the lab frame $S$, the frame fixed on the 2-D substrate $S'$ and the body frame of one particular molecule $S''$.}
\end{figure}

The sample used in RCT consists of multiple copies of partially oriented molecules. Specifically, the molecules adsorb to a 2-D substrate such that a particular axis within the molecules aligns with the substrate normal. The molecular orientations are limited to rotations about the particular body axis by angles uniformly drawn from $[0, 2\pi)$. Let $S''$ be the body frame of one particular molecule, where the $\mathbf{z''}$-axis coincides with its body rotation axis. We further define another reference frame $S'$ fixed on the 2-D substrate such that the $\mathbf{x'}$-axis coincides with the tilt axis of the substrate and the $\mathbf{z'}$-axis aligns with the substrate normal. In the following, we also assume that the $\mathbf{x}$-axis of the lab frame is parallel to the $\mathbf{x'}$-axis. After specifying these reference frames, we define the substrate tilt angle $\theta$ as the angle between the $\mathbf{z}$- and $\mathbf{z'}$-axes. The rotation angle $\alpha$ of the particular molecule with respect to its body rotation axis is defined as the angle between the $\mathbf{x'}$- and $\mathbf{x''}$-axes. The relationships between the reference frames are shown in Figure~\ref{fig:coordinates}.

Let $f(x'', y'', z'')$ be the 3-D Coulomb potential of the particular molecule in its own body frame $S''$. Hereafter, we refer to $f$ as the structure of the molecule. From the geometries shown in Figure~\ref{fig:coordinates}, the coordinate transformation between $S$ and $S''$ is given by
\begin{align}
\label{eq:coor_trans}
\mathbf{r} &= 
\begin{bmatrix}
	\cos\alpha & -\sin\alpha & 0 \\
	\cos\theta\sin\alpha & \cos\theta\cos\alpha & -\sin\theta \\
	\sin\theta\sin\alpha & \sin\theta\cos\alpha & \cos\theta
\end{bmatrix}
\mathbf{r''} + \mathbf{t} \nonumber \\
&= R^\theta_\alpha \mathbf{r''} + \mathbf{t},
\end{align}
where $\mathbf{r} = [x, y, z]^T$, $\mathbf{r''} = [x'', y'', z'']^T$, $R^\theta_\alpha$ is the rotation matrix that aligns the axes of $S$ with the axes of $S''$, and $\mathbf{t} = [t_x, t_y, t_z]^T$ is the vector pointing from the origin of $S$ to the origin of $S''$. We can therefore express the molecular structure in the lab frame $S$ by $f((R^\theta_\alpha)^T(\mathbf{r}-\mathbf{t}))$, and its tomographic projection along the $\mathbf{z}$-axis is given by $\mathcal{I}^\theta_\alpha(x - t_x, y - t_y)$, where
\begin{equation}
\label{eq:proj_img}
\mathcal{I}^\theta_\alpha(x, y) = \int_{-\infty}^{\infty} f((R^\theta_\alpha)^T\mathbf{r})~dz.
\end{equation}

Taking the 2-D Fourier transform on both sides of \eqref{eq:proj_img}, with the Fourier slice theorem, we get
\begin{equation}
\label{slice_thm}
\hat{\mathcal{I}}^\theta_\alpha(k_x, k_y) = \hat{f}((R^\theta_\alpha)^T [k_x, k_y, 0]^T),
\end{equation}
where $\hat{f}(k_x, k_y, k_z)$ denotes the 3-D Fourier transform of $f(x, y, z)$. As a result, a projection image contains the same information as the central slice of the 3-D Fourier transform that is perpendicular to the direction of projection. Since the molecular orientations are limited to in-plane rotations on the \mbox{2-D} substrate, which is itself tilted by an angle $\theta$, the corresponding Fourier slices fill the whole 3-D Fourier space except for the region within a double cone, whose axis coincides with the body rotation axis of the molecules. The double cone has an opening angle $2\theta$ and the region within the missing cone represents the inaccessible information of the molecular structure in the setting of RCT.

\subsection{Micrograph simulation}
\label{sec:micrograph}
Before discussing our model for simulating micrographs, we first consider the computation of the molecular projection images. Let $F$ be the discretization of the molecular structure $f$ that is defined on a cubic grid $(x, y, z) \in \{-2r, \dots, 2r\}^3$ by \begin{equation}
F(x, y, z) = f(x/\xi, y/\xi, z/\xi).
\end{equation}
The integer $r$ represents the radius of a spherical support such that $F(x, y, z)$ is negligible for $(x^2 + y^2 + z^2)^{1/2} \geq r$. In addition, we define the discretization of the molecular projection $\mathcal{I} ^\theta_\alpha$ by
\begin{equation}
I^\theta_\alpha(x, y) = \mathcal{I} ^\theta_\alpha(x/\xi, y/\xi), 
\end{equation}
where $(x, y) \in \{-2r, \dots, 2r\}^2$, and it immediately follows that $I^\theta_\alpha$ has a circular support of radius $r$. From the Fourier slice theorem, we can compute the discrete Fourier transform (DFT) of $I^\theta_\alpha$ from the DFT of $F$ by
\begin{equation}
\hat{I}^\theta_\alpha(k_x, k_y) = \hat{F}((R^\theta_\alpha)^T [k_x, k_y, 0]^T),
\end{equation}
where $(k_x, k_y) \in \{-2r, \dots, 2r\}^2$. To reduce the interpolation error, we use the FINUFFT package~\cite{Barnett2019,Barnett2020} to evaluate $\hat{F}$ on the non-uniform grid points. Finally, we obtain the molecular projections $I^\theta_\alpha$ by the inverse DFT of $\hat{I}^\theta_\alpha$.

\begin{figure}
\label{fig:micrograph}
\includegraphics[scale=0.75, trim=3.5cm 1.5cm 3cm 1.5cm, clip=true]{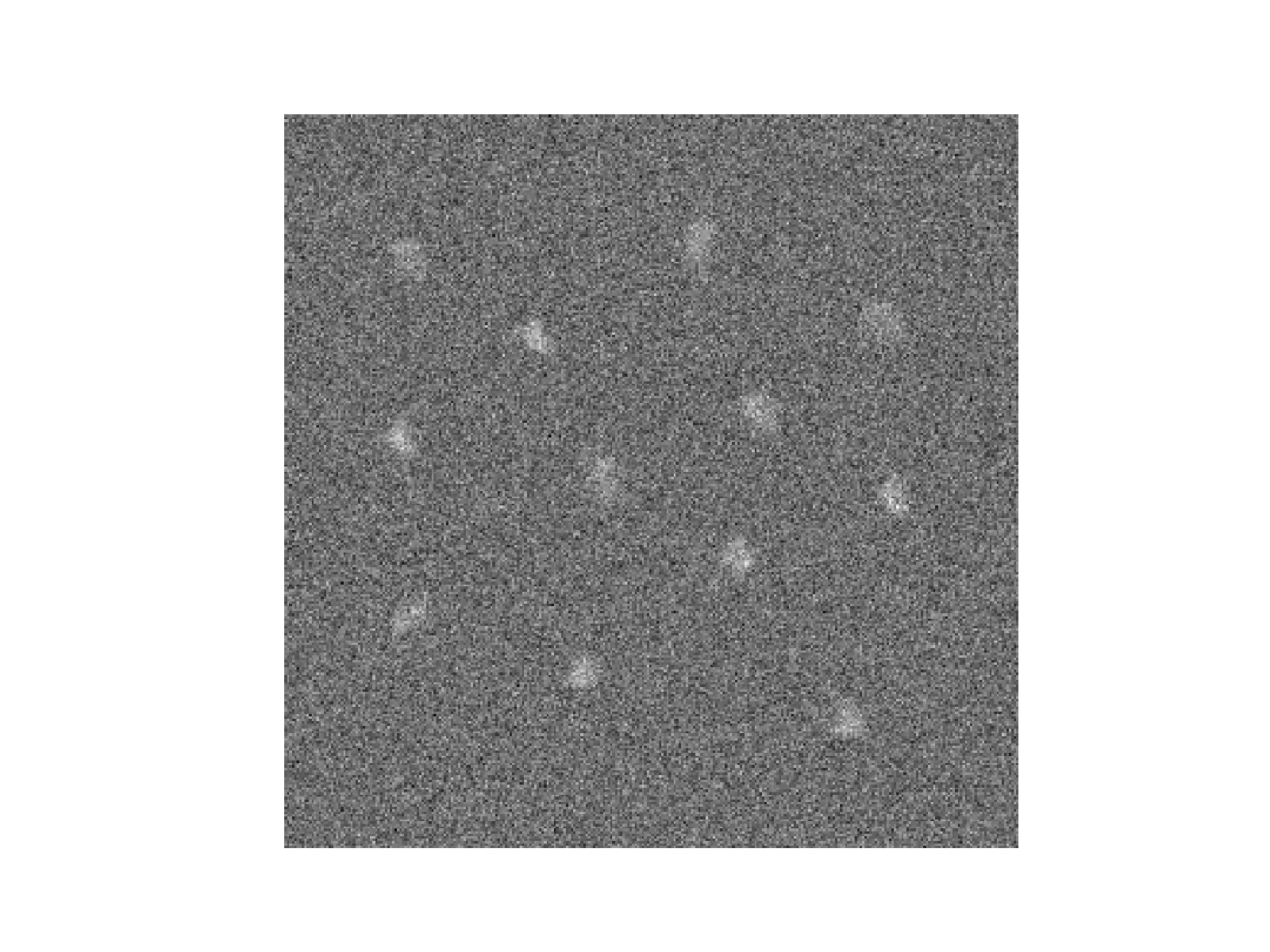}
\caption{A sample micrograph with SNR = 1.}
\end{figure}

We simulate the micrographs measured in a RCT experiment at the substrate tilt angle $\theta$ by
\begin{equation}
M(x_i, y_i) = \sum_{j=1}^{n_p} I^\theta_{\alpha_j}(x_i - t_{x_j}, y_i - t_{y_j}) + \varepsilon(x_i, y_i),
\end{equation}
where $(x_i, y_i) \in \{ -\lfloor m/2 \rfloor, \dots, \lceil m/2 - 1 \rceil \}^2$, $n_p$ is the number of molecular projections in the micrograph, $\alpha_j$ is the in-plane rotation of the $j^\mathrm{th}$ molecule that is uniformly drawn from $[0, 2\pi)$, $(t_{x_j}, t_{y_j}) \in \{ -\lfloor m/2 \rfloor + r, \dots, \lceil m/2 - 1 \rceil - r \}^2$ is the center of the tomographic projection of the $j^\mathrm{th}$ molecule, and $\varepsilon(x_i, y_i)$ is i.i.d. Gaussian noise with zero mean and variance $\sigma^2$. For a reason that will be clear in Section~\ref{sec:autocorr}, we further assume that 
\begin{equation}
\label{eq:separation}
((t_{x_j} - t_{x_k})^2 + (t_{y_j} - t_{y_k})^2)^{1/2} > 4r\quad\mathrm{for}~j \neq k 
\end{equation}
such that the molecular projections are well separated in the micrographs. Figure~\ref{fig:micrograph} shows a sample micrograph with SNR = 1. We define SNR as the ratio of the mean squared pixel values of molecular projections to the noise variance. Specifically,
\begin{equation}
\mathrm{SNR} = \frac{1}{2\pi}\int_0^{2\pi} d\alpha~\frac{1}{\pi r^2}\sum_{x_i^2 + y_i^2 < r^2} |I^\theta_\alpha(x_i, y_i)|^2 \bigg/ \sigma^2.
\end{equation}

\subsection{Auto\-correlation analysis}
\label{sec:autocorr}
The standard data processing pipelines in single-particle cryo-EM start with the step of particle picking to locate the molecular projections in the noisy micrographs, which is equivalent to determining the 2-D vector $[t_{x_j}, t_{y_j}]^T$ for each molecular projection. This task, however, becomes challenging when the noise level is high. An alternative is to extract from the data quantities that are invariant to the 2-D translations of molecular projections in the micrographs. We achieve this through the approach of autocorrelation analysis. 

Consider an $n \times n$ image $g(\mathbf{x})$. We define its autocorrelation function of order $q = 1, 2, \dots$ for any 2-D translations $\mathbf{x_1}, \dots, \mathbf{x_{q-1}} \in \mathbb{Z}^2$ by
\begin{equation}
a_g^q(\mathbf{x_1}, \dots, \mathbf{x_{q-1}}) 
= \frac{1}{n^2} \sum_{\mathbf{x}} g(\mathbf{x})g(\mathbf{x}+\mathbf{x_1}) \cdots g(\mathbf{x}+\mathbf{x_{q-1}}),
\end{equation}
where $\mathbf{x} \in \{ -\lfloor n/2 \rfloor, \dots, \lceil n/2 - 1 \rceil \}^2$ and $g(\mathbf{x})$ is zero-padded for arguments out of the range. In the context of this study, we set $n = m$ when $g$ represents a micrograph $M$ and $n = 4r+1$ when $g$ represents a molecular projection $I^\theta_\alpha$. 

Under the assumption that the molecular projections are well separated, as in \eqref{eq:separation}, the autocorrelations of a micrograph with \mbox{2-D} translations $\mathbf{x_1}, \dots, \mathbf{x_{q-1}}$, where $|\mathbf{x_1}|, \dots, |\mathbf{x_{q-1}}| \leq 2r$, are insensitive to the locations of molecular projections in the micrograph. As a result, the micrograph autocorrelations can be directly related to the autocorrelations of molecular projections, which provide information about the molecular structure. 

In this work, we consider the micrograph autocorrelations up to the third order. Under the additional assumption that the density of molecular projections \mbox{$\gamma = n_p(4r+1)^2/m^2$} is fixed, it is straightforward to show that (see for example in \citeasnoun{Bendory2018})
\begin{align}
\label{eq:aM1}
\mathbb{E}\{a^1_M\} &= \gamma~\langle a^1_{I^\theta_\alpha} \rangle_\alpha \\
\label{eq:aM2}
\mathbb{E}\{a^2_M(\mathbf{x_1})\} &= 
	\gamma~\langle a^2_{I^\theta_\alpha}({\mathbf{x_1}}) \rangle_\alpha + \sigma^2 \delta({\mathbf{x_1}}) \\
\label{eq:aM3}
\mathbb{E}\{a^3_M(\mathbf{x_1}, \mathbf{x_2})\} &= 
	\gamma~\langle a^3_{I^\theta_\alpha}({\mathbf{x_1} + \mathbf{x_2}}) \rangle_\alpha \nonumber \\
	&\quad+~\gamma~\langle a^1_{I^\theta_\alpha} \rangle_\alpha~\sigma^2 \big( \delta({\mathbf{x_1}}) 
	+ \delta({\mathbf{x_2}}) + \delta({\mathbf{x_1} - \mathbf{x_2}}) \big)
\end{align}
for any fixed level of noise and $|\mathbf{x_1}|, |\mathbf{x_2}| \leq 2r$. Here $\mathbb{E}\{\cdot\}$ represents the expectation over the distributions of the random Gaussian noise and the in-plane rotations of molecules, and $\langle\cdot\rangle_\alpha$ denotes the angular average over $\alpha \in [0, 2\pi)$. The delta functions, defined by $\delta(0) = 1$ and $\delta({\mathbf{x}} \neq 0) = 0$, are due to the autocorrelations of the random Gaussian noise. 

We estimate the expectations in \eqref{eq:aM1}-\eqref{eq:aM3} by averaging autocorrelations computed from many micrographs. In practice, $\sigma^2$ and $\gamma~\langle a^1_{I^\theta_\alpha} \rangle_\alpha$ can be estimated from the micrographs: $\sigma^2$ can be estimated by the variance of micrograph pixel values in the low SNR regime; $\gamma~\langle a^1_{I^\theta_\alpha} \rangle_\alpha$ can be estimated by the empirical mean of micrographs. As a result, we can estimate the autocorrelations $\langle a^1_{I^\theta_\alpha} \rangle_\alpha$, $\langle a^2_{I^\theta_\alpha}({\mathbf{x_1}}) \rangle_\alpha$ and $\langle a^3_{I^\theta_\alpha}({\mathbf{x_1}, \mathbf{x_2}}) \rangle_\alpha$ up to the constant factor $\gamma$. For simplicity, we assume that $\sigma^2$ and $\gamma$ are known to us. 

\subsection{Regularized optimization}
\label{sec:opt}
In this section, we design an optimization problem to reconstruct the molecular structure $F$ from the autocorrelations computed from micrographs. We start by expressing $F$ in a non-redundant representation. Recall that $F$ is defined on a cubic grid of size $4r+1$ and has a spherical support of radius $r$. We represent $F$ by a vector $\mathbf{u}$ of length $n_r$, where $n_r$ denotes the number of voxels within the support. Furthermore, we define the linear operator $\mathcal{A}$ that maps $\mathbf{u}$ to $F$ by $F = \mathcal{A}\mathbf{u}$.

Our optimization problem estimates $\mathbf{u}$ by fitting the rotationally averaged $3^\mathrm{rd}$ order autocorrelation $\langle a^3_{I^\theta_\alpha}({\mathbf{x_1}, \mathbf{x_2}}) \rangle_\alpha$. As will be seen later, $\langle a^1_{I^\theta_\alpha} \rangle_\alpha$ is used to generate the initial guess for $\mathbf{u}$, and $\langle a^2_{I^\theta_\alpha}({\mathbf{x_1}}) \rangle_\alpha$ is used to build the regularizer in the optimization. For computational efficiency, we construct the cost function with the DFT of $\langle a^3_{I^\theta_\alpha}({\mathbf{x_1}, \mathbf{x_2}}) \rangle_\alpha$, where
\begin{align}
\label{eq:s3}
s_F^3({\mathbf{k_1}, \mathbf{k_2}}) &= \mathcal{F} \{ \langle a^3_{I^\theta_\alpha}({\mathbf{x_1}, \mathbf{x_2}}) \rangle_\alpha \} ({\mathbf{k_1}, \mathbf{k_2}}) \nonumber \\
&= \frac{1}{2\pi}\int_0^{2\pi} d\alpha~\hat{I^\theta_{\alpha}}({\mathbf{k_1}}) \hat{I^\theta_{\alpha}}({\mathbf{k_2}}) \hat{I^\theta_{\alpha}}^\ast({\mathbf{k_1+k_2}}) \nonumber \\
&\approx \frac{1}{n_\alpha} \sum_{i=0}^{n_\alpha-1} \hat{I^\theta_{\alpha_i}}({\mathbf{k_1}}) \hat{I^\theta_{\alpha_i}}({\mathbf{k_2}}) \hat{I^\theta_{\alpha_i}}^\ast({\mathbf{k_1+k_2}}),
\end{align}
where $*$ denotes the complex conjugate. In the last step, we replace the integration with a discrete sum over $n_\alpha$ samples, where $\alpha_i = 2\pi i/n_\alpha$. 

The triple product in \eqref{eq:s3} is the Fourier transform of the $3^\mathrm{rd}$ order autocorrelation $a^3_{I^\theta_{\alpha_i}}({\mathbf{x_1}, \mathbf{x_2}})$, also known as the bispectrum~\cite{Tukey1953}. Its applications in signal processing can be seen, for instance, in \citeasnoun{Sadler1992} and \citeasnoun{Bendory2017}. Since we assume that the information of the molecular projections is preserved only up to the Nyquist frequency due to noise, we only consider spatial frequencies $(\mathbf{k_1}, \mathbf{k_2}) \in \mathcal{V}$, where $\mathcal{V} = \{(\mathbf{k_1}, \mathbf{k_2}): |\mathbf{k_1}|, |\mathbf{k_2}|, |\mathbf{k_1+k_2}| < 2r \}$. Let $\tilde{s}_F^3({\mathbf{k_1}, \mathbf{k_2}})$ be the DFT of the estimation of $\langle a^3_{I^\theta_\alpha}({\mathbf{x_1}, \mathbf{x_2}}) \rangle_\alpha$ from data. We can hence express the sum of least-square errors by $\sum_{(\mathbf{k_1}, \mathbf{k_2}) \in \mathcal{V}} \big| s_F^3({\mathbf{k_1}, \mathbf{k_2}}) - \tilde{s}_F^3({\mathbf{k_1}, \mathbf{k_2}}) \big|^2$.

As discussed in Section~\ref{sec:rct}, there exists a double-cone region in the Fourier space that cannot be probed in RCT. Therefore, our reconstruction problem is ill-posed in nature, and we must include a regularization term in the cost function to incorporate some prior knowledge of the true solution. Our regularization enforces the smoothness assumption on $F$ and has the form of the weighted sum of squares: $\sum_{\mathbf{q}} |\hat{F}({\mathbf{q}})|^2/\tau({\mathbf{q}})^2$, where $\mathbf{q} \in \{-2r, \dots, 2r\}^3$. This regularization is related to the Gaussian prior described in \citeasnoun{Scheres2012} in that we expect the scale parameters $\tau({\mathbf{q}})^2$ to act as a low-pass filter to reduce high-frequency noise while still preserve some high-resolution features of the molecule. 

We estimate the values of $\tau({\mathbf{q}})$ based on the observation that the structure factors of proteins obey Wilson statistics~\cite{Wilson1949}. To be more precise, the structure factors within each resolution shell follow the complex normal distribution with mean zero and variance estimated from the mean intensity in the resolution shell~\cite{French1978}. Taking the DFT of $\langle a^2_{I^\theta_\alpha}({\mathbf{x_1}}) \rangle_\alpha$, we obtain
\begin{align}
s_F^2({\mathbf{k_1}}) &= \mathcal{F}\{\langle a^2_{I^\theta_\alpha}({\mathbf{x_1}}) 
	\rangle_\alpha\} ({\mathbf{k_1}})
	= \frac{1}{2\pi}\int_0^{2\pi} d\alpha~\hat{I^\theta_{\alpha}}({\mathbf{k_1}}) 
	\hat{I^\theta_{\alpha}} ^\ast({\mathbf{k_1}}) \nonumber \\
	&= \frac{1}{2\pi}\int_0^{2\pi} d\alpha~|\hat{F}((R^\theta_\alpha)^T [k_{1x}, k_{1y}, 0]^T)|^2,
\end{align}
where $\mathbf{k_1} \in \{-2r, \dots, 2r\}^2$ and we only consider spatial frequencies within the Nyquist frequency, that is, $|\mathbf{k_1}| < 2r$. Since
\begin{align}
(R^\theta_\alpha)^T \begin{bmatrix} k_{1x} \\ k_{1y} \\ 0 \end{bmatrix} =
	\begin{bmatrix} 
		k_{1x}\cos\alpha + k_{1y}\cos\theta\sin\alpha \\
		- k_{1x}\sin\alpha + k_{1y}\cos\theta\cos\alpha \\
		- k_{1y}\sin\theta 
	\end{bmatrix},
\end{align}
we can see that $s_F^2({\mathbf{k_1}})$ is the mean intensity over a circle that is perpendicular to the body rotation axis of the molecule and has radius $(k_{1x}^2 + k_{1y}^2\cos^2\theta)^{1/2}$. Therefore, with appropriate weights, the average of $s_F^2({\mathbf{k_1}})$ for all $\mathbf{k_1}$ that fall into the same annulus $q_\mathrm{min} < |\mathbf{k_1}| < q_\mathrm{max}$ gives the mean intensity within the resolution shell $q_\mathrm{min} < |\mathbf{q}| < q_\mathrm{max}$, excluding the spherical caps that lie in the missing-cone region. We represent this weighted average by $\tau_{|\mathbf{q}|}^2$, whose values are in practice computed from $\tilde{s}_F^2({\mathbf{k_1}})$, the DFT of the estimation of $\langle a^2_{I^\theta_\alpha}({\mathbf{x_1}}) \rangle_\alpha$ from data. 

In addition to the scale parameters $\tau_{|\mathbf{q}|}^2$ for $|\mathbf{q}| < 2r$, it is helpful to have regularization outside the Nyquist frequency to limit high-frequency noise. We choose $\tau(\mathbf{q}) = |\mathbf{q}|^{-1}$ for $|\mathbf{q}| \geq 2r$. This choice is based on the identity
\begin{equation}
\int_{\mathbb{R}^3} |\nabla f|^2~d^3\mathbf{x} = \int_{\mathbb{R}^3} |\mathbf{q}|^2 |\hat{f}(\mathbf{q})|^2~d^3\mathbf{q}
\end{equation}
such that one can minimize the sum of gradient squares by minimizing $\int_{\mathbb{R}^3} |\mathbf{q}|^2 |\hat{f}(\mathbf{q})|^2~d^3\mathbf{q}$. Finally, we define the cost function by
\begin{align}
\label{eq:cost}
C({\mathbf{u}}) &= \sum_{({\mathbf{k_1}, \mathbf{k_2}}) \in \mathcal{V}} \big| s_F^3({\mathbf{k_1}, \mathbf{k_2}}) 
	- \tilde{s}_F^3({\mathbf{k_1}, \mathbf{k_2}}) \big|^2 \nonumber \\
	&+ \lambda \left( \sum_{|\mathbf{q}| < 2r} \frac{|\hat{F}(\mathbf{q})|^2}{\tau_{|\mathbf{q}|}^2} + \beta \sum_{|\mathbf{q}| \geq 2r} |\mathbf{q}|^2 |\hat{F}(\mathbf{q})|^2 \right),
\end{align}
where ${\mathbf{u}}$ is the non-redundant representation of $F$, $\lambda$ denotes the regularization parameter, and we compute the scale factor $\beta$ such that the two curves $\tau_{|\mathbf{q}|}^2$ and $|\mathbf{q}|^{-2}$ attain the same value at $|\mathbf{q}| = 2r$. We have also attempted optimization with $\sum_{\mathbf{q}} |\mathbf{q}|^2 |\hat{F}(\mathbf{q})|^2$ as the only regularization term, but the quality of the reconstruction appears to be inferior (not shown in this study).

\section{Results}
\label{sec:results}
\subsection{Reconstruction at different substrate tilts}
In this section, we explore the effect of the missing-cone region on the quality of reconstruction by considering micrographs measured at different substrate tilt angles $\theta = 60^\circ, 35^\circ$ and $10^\circ$. The molecule used in our simulation is Bovine Pancreatic Trypsin Inhibitor (BPTI), which has size of 35~\AA~and weight of $6.5~\mathrm{kDa}$. This molecular size is substantially below the limit ($40~\mathrm{kDa}$) believed to be attainable by single-particle cryo-EM~\cite{Henderson1995}, and our model structure was determined using X-ray crystallography. 

We generate the discrete molecular structure $F$ from the PDB entry 1QLQ~\cite{BPTI_pdb} using the UCSF Chimera software~\cite{Chimera} at a resolution of 5~\AA. The resulting contrast has a spherical support of radius $r = 15$ voxels, and is further zero-padded to be a cubic grid of size $61$. From the discrete contrast $F$, we simulate the micrographs as described in Section~\ref{sec:micrograph}. To obtain the baseline results on the effect of the missing cone region, we consider the idealistic scenario that the in-plane rotation of the $j^{\mathrm{th}}$ molecule is given by $\alpha_j = 2\pi j/n_p$, $j \in \{1, \dots, n_p\}$, and the noise variance $\sigma^2 = 0$. By setting the micrograph length $m = 4096$ pixels and the number of molecules $n_p = 400$, we only simulate one micrograph at each given value of the substrate tilt angle.

\begin{figure}
\label{fig:rad_ave}
\includegraphics[scale=0.6, trim=0cm 0cm 0cm 0cm, clip=true]{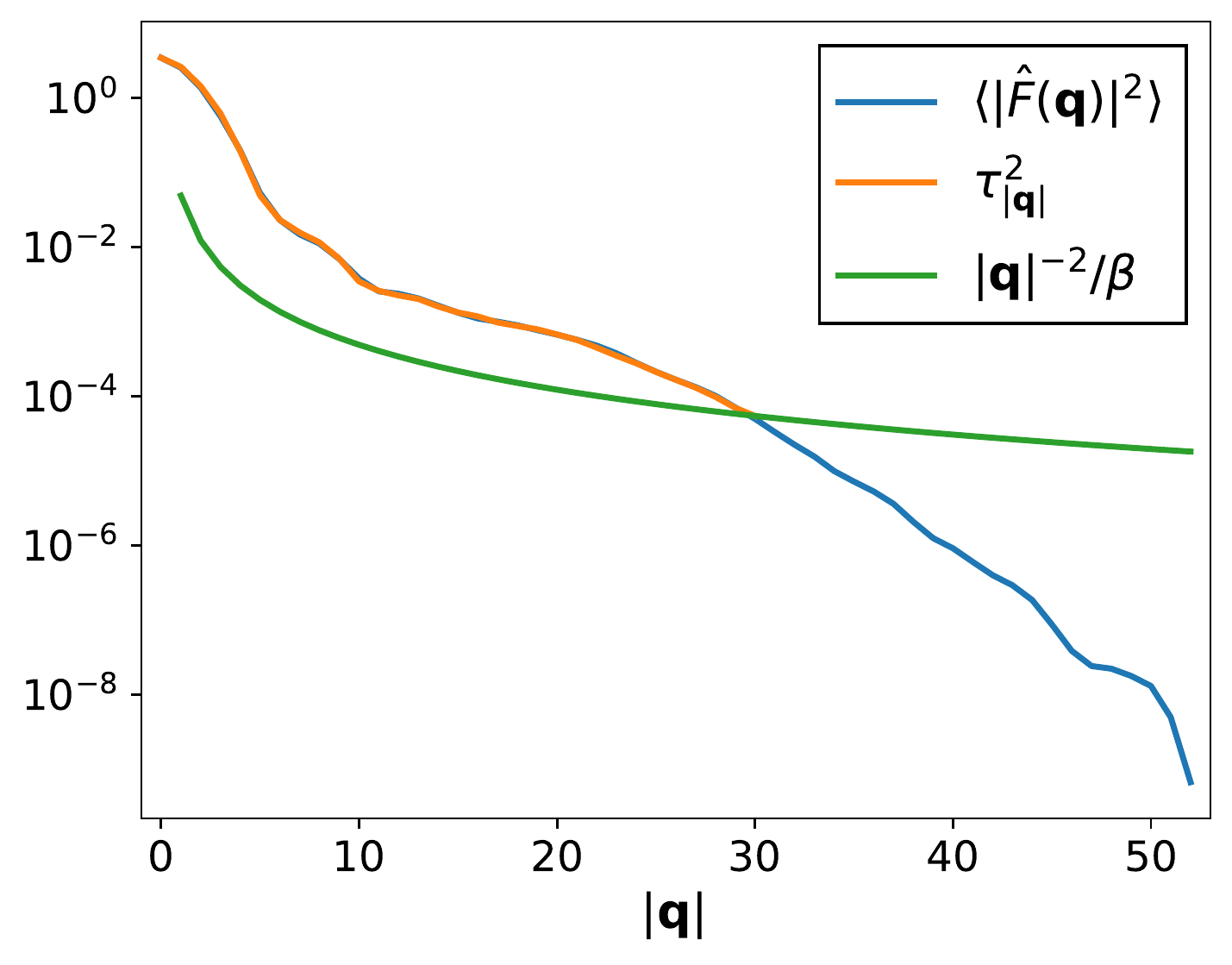}
\caption{The comparison of the mean intensities $\langle |\hat{F}(\mathbf{q})|^2 \rangle$ and the scale parameters $\tau_{|\mathbf{q}|}^2$ and $|{\mathbf{q}}|^{-2}/\beta$ for the BPTI molecule at the substate tilt angle $\theta = 60^\circ$.}
\end{figure}

From the simulated micrographs, we compute the rotationally averaged autocorrelations of molecular projections and the values of $\tilde{s}_F^3({\mathbf{k_1}, \mathbf{k_2}})$ and $\tau_{|\mathbf{q}|}^2$. Figure~\ref{fig:rad_ave} shows the comparison of the mean intensities $\langle |\hat{F}(\mathbf{q})|^2 \rangle$ and the scale parameters $\tau_{|\mathbf{q}|}^2$ and $|{\mathbf{q}}|^{-2}/\beta$ for $\theta = 60^\circ$. We first see that $\tau_{|\mathbf{q}|}^2$ provides a good estimate for $\langle |\hat{F}(\mathbf{q})|^2 \rangle$ up to the Nyquist frequency. On the other hand, the scale parameter $|{\mathbf{q}}|^{-2}/\beta$ is substantially greater than $\langle |\hat{F}(\mathbf{q})|^2 \rangle$ outside the Nyquist frequency, which may inevitably preserve some high-resolution noise in the reconstruction. 

We use the BFGS algorithm in the tensorflow software library~\cite{Tensorflow} to minimize the cost function \eqref{eq:cost} over a set of regularization parameters $\lambda = 10^{-2}, 10^{-1}, \dots, 10^7$. In order to speed up the optimization, we initialize $\mathbf{u}$ from the discretization of a 3-D Gaussian profile, whose variance is set as $r^2$. We further rescale $\mathbf{u}$ such that the sum of the corresponding $F$ is equal to the estimate for $(4r+1)^2\langle a^1_{I^\theta_\alpha} \rangle_\alpha$ from data. From the converged solutions, we choose the optimal value of $\lambda$ using the L-curve method~\cite{Hansen1992}. Our reconstructed structures are the estimates for $F$ with these optimal values of $\lambda$.

\begin{figure}
\label{fig:tilts}
\includegraphics[scale=0.42, trim=0cm 0cm 0cm 0cm, clip=true]{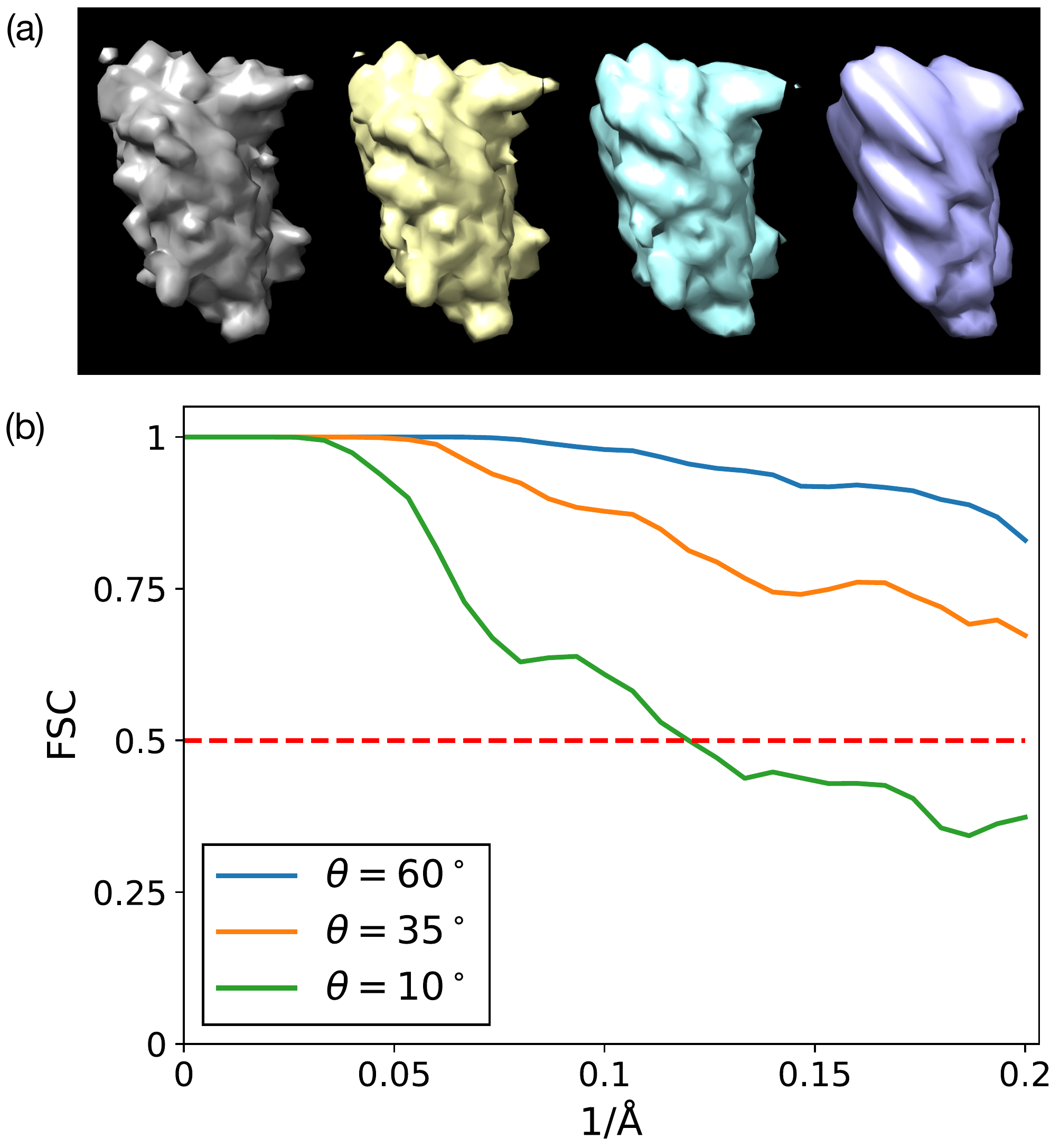}
\caption{(a) The reconstructed BPTI structures from noiseless micrographs at different sample tilt angles: $\theta = 60^\circ$ (yellow), $\theta = 35^\circ$ (cyan) and $\theta = 10^\circ$ (purple). The grey one is the ground truth used to simulate the micrographs. (b) The FSC of the reconstructed structures with the ground truth.}
\end{figure}

Figure~\ref{fig:tilts}(a) shows the comparison of the reconstructed BPTI structures with the ground truth used to simulate the micrographs. As expected, the visual quality of the reconstructions degrades when the sample tilt angle $\theta$ decreases, which results in a larger missing-data region. To assess the reconstructions in more detail, we plot the Fourier shell correlation (FSC)~\cite{Harauz1986} of the reconstructed structures with the ground truth in Figure~\ref{fig:tilts}(b). Although the reconstruction at $\theta=35^\circ$ correlates to the ground truth worse than the one at $\theta=60^\circ$, both of them have the same resolution as the ground truth (5~\AA) according to the FSC = 0.5 criterion. Using the same criterion, the resolution of the reconstruction at $\theta=10^\circ$ is 8.3~\AA.

\subsection{Reconstruction from noisy micrographs}
After having the baseline results for reconstructions from noiseless micrographs, we turn to test our approach on noisy micrographs. At the sample tilt angle $\theta = 60^\circ$, we simulate 500 micrographs of size $m = 4096$ using the same discrete contrast $F$ for BPTI. We adjust the noise level such that the micrographs have SNR = 1. By maximizing the density of molecular projections while still preserving the requirement of well separation~\eqref{eq:separation}, the resulting micrographs contain $1.4\times10^6$ molecular projections in total. 

\begin{figure}
\label{fig:BPTI_snr1}
\includegraphics[scale=0.35, trim=0cm 0cm 0cm 0cm, clip=true]{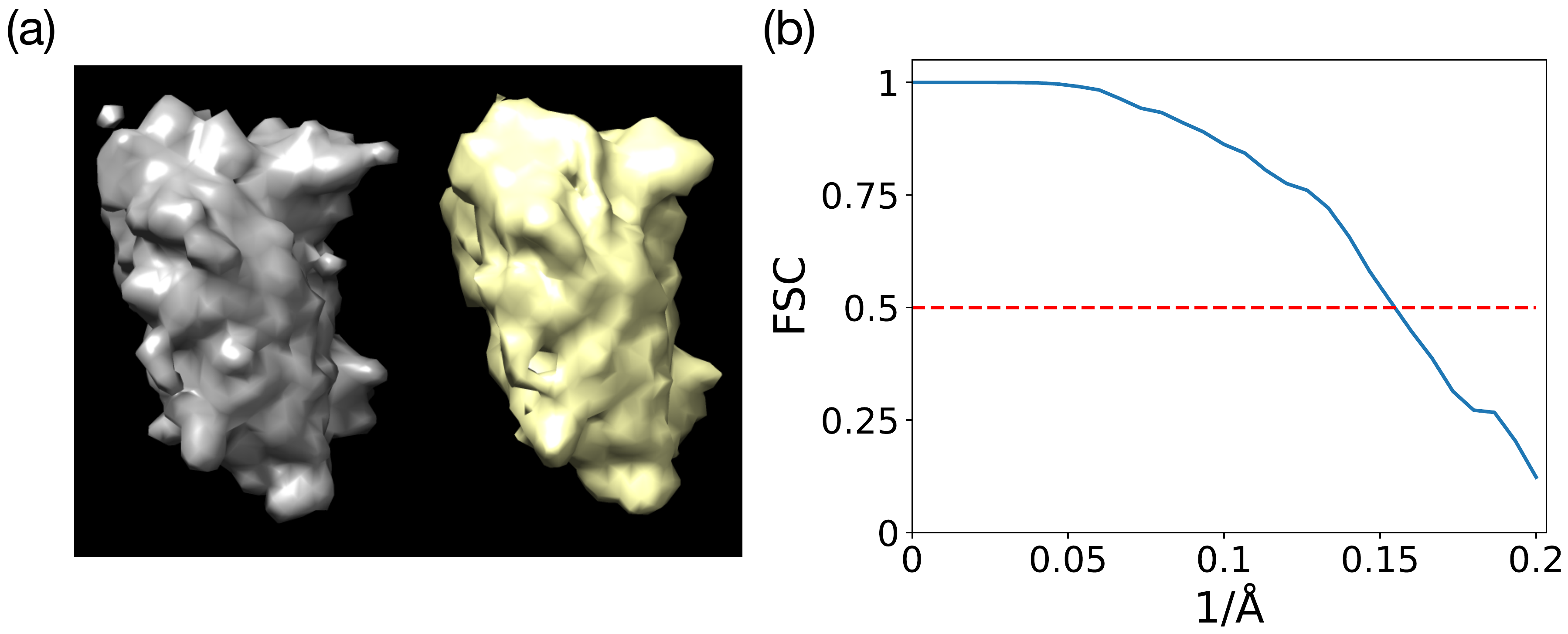}
\caption{(a) The reconstructed BPTI structure (yellow) from noisy micrographs with SNR = 1 at the sample tilt angle $\theta = 60^\circ$. The ground truth is rendered in grey. (b) The FSC of the reconstructed structure with the ground truth.}
\end{figure}

From the noisy micrographs, we compute the estimates for the rotationally averaged autocorrelations of molecular projections. Figure~\ref{fig:BPTI_snr1}(a) shows the reconstruction from these estimates along with the ground truth. The negative effect of noise on the quality of the reconstruction can best be seen by comparing this reconstruction with its counterpart in Figure~\ref{fig:tilts}(a). As plotted in Figure~\ref{fig:BPTI_snr1}(b), we determine the resolution of this reconstructed structure to be 6.5~\AA~using the FSC = 0.5 criterion. 

\begin{figure}
\label{fig:myoglobin_snr1}
\includegraphics[scale=0.35, trim=0cm 0cm 0cm 0cm, clip=true]{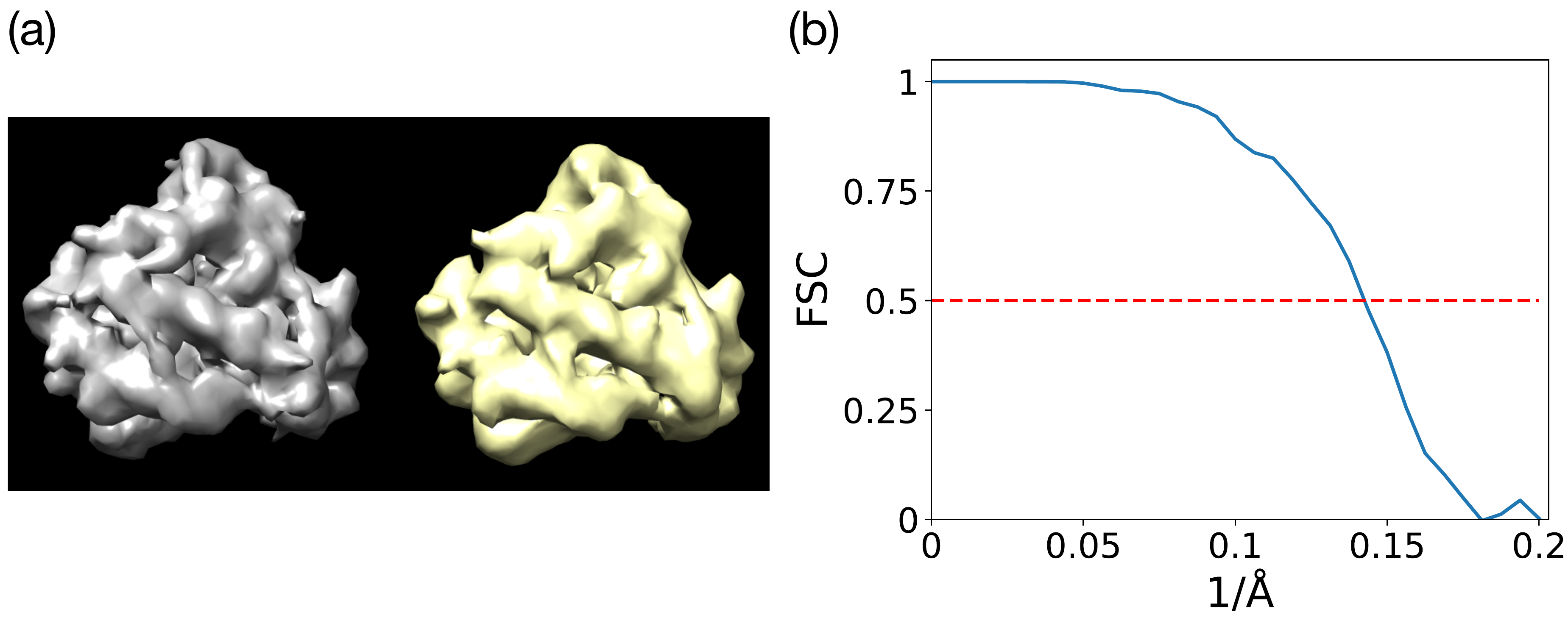}
\caption{(a) The reconstructed myoglobin structure (yellow) from noisy micrographs with SNR = 1 at the sample tilt angle $\theta = 60^\circ$. The ground truth is rendered in grey. (b) The FSC of the reconstructed structure with the ground truth.}
\end{figure}

To demonstrate that our approach applies to other biological molecules, we test our approach on another dataset simulated from the myoglobin molecule, which has size of 40~\AA~and weight of 17.8 kDa. We generate the discrete molecular structure $F$ for myoglobin from the PDB entry 1MBN~\cite{myoglobin_pdb} using the UCSF Chimera software at a resolution of 5~\AA. The resulting contrast has a spherical support of radius $r = 16$ voxels, and is further zero-padded to be a cubic grid of size $65$. At the sample tilt angle $\theta = 60^\circ$, we generate 500 micrographs of size $m = 4096$ from $F$. The number of molecular projections in these micrographs totals $1.2\times 10^6$, and we also set SNR = 1 for the micrographs. The reconstructed myoglobin structure from the noisy micrographs is shown in Figure~\ref{fig:myoglobin_snr1}(a) along with the ground truth. We can see that our reconstruction recovers most of the main features of the ground truth. We plot the FSC of our reconstruction with the ground truth in Figure~\ref{fig:myoglobin_snr1}(b), and we determine the resolution of the reconstruction to be 7.0~\AA~according to the FSC = 0.5 criterion.

\section{Discussion}
\label{sec:discussion}
In this paper, we present a method to reconstruct the \mbox{3-D} molecular structure from data collected at just one sample tilt angle in RCT. Our method reduces data to quantities that are invariant to the 2-D positions of molecular projections in the micrographs, which removes the need for particle picking when analyzing data. In order to address the missing data in the double-cone region of the molecule's Fourier transform, we design a regularized optimization problem to reconstruct the molecular structure by fitting the autocorrelations estimated from micrographs. Our numerical studies illustrate the effect of the missing-cone region on the quality of reconstruction. In addition, we demonstrate structure reconstruction from the autocorrelations computed from noisy micrographs. Since the accuracy of the autocorrelation estimates can be improved by averaging many more micrographs, our results show promise of applying autocorrelation analysis to reconstruct the structures of small biological molecules in the setting of RCT. 

A few issues still stand in the way of applying our approach to real RCT data. In Section~\ref{sec:img_form}, we make the assumption that the point spread function is a 2-D Dirac delta function to ignore its effect. In reality, however, we may have to consider a varying point spread function with respect to the locations on the detector because different regions of the tilted specimen are exposed to the electron beam with different defocus values. Another challenge arises when the noise is colored. In that case, the expectations of products of noise at different pixels are not zero. It will require a more sophisticated model for the noise power spectrum instead of a single parameter $\sigma^2$. Furthermore, structure heterogeneity of the target molecule will be another test for our approach.

Additionally, we assume that the molecular projections are well separated in the micrographs. This assumption enables us to directly relate the micrograph autocorrelations to the autocorrelations of molecular projections. However, it is preferable in practice to have the molecular projections densely packed in micrographs to maximize the available structural information within limited data collection time. We expect to remove this assumption by considering the cross correlations between neighboring molecular projections. A similar idea was recently demonstrated in~\citeasnoun{Lan2020} for the simplified 1-D model. 

Another practical concern is the amount of required data. As a proof of concept, we reconstruct the molecular structures from simulated micrographs with SNR = 1. For small biological molecules that challenges particle picking, we expect the SNR of the micrographs to be much lower. Since our approach uses autocorrelations up to the $3^{\mathrm{rd}}$ order, the sample complexity would scale as $\mathrm{SNR}^{-3}$. This means that we will need $10^3$ times more molecules to estimate the autocorrelations with similar accuracy when the SNR drops from 1 to 0.1. Although densely packing the molecular projections in micrographs helps improve the SNR of the estimated autocorrelations, it would be beneficial to investigate methods to denoise the autocorrelations. 

In the long run, we would like to extend the approach described here to real cryo-EM data to reconstruct high-resolution structures directly from micrographs, without being restricted to molecules which have a preferred orientation on their substrate.

\section{Acknowledgements}
TYL and AS were supported in part by AFOSR Awards FA9550-17-1-0291 and FA9550-20-1-0266, the Simons Foundation Math+X Investigator Award, the Moore Foundation Data-Driven Discovery Investigator Award, NSF BIGDATA Award IIS-1837992, NSF Award DMS-2009753, and NIH/NIGMS Award R01GM136780-01. We would like to thank Tamir Bendory, Joe Kileel, Eitan Levin and Nicholas Marshall for productive discussions. 


\appendix
\section{Computational Details}
\label{sec:comp-details}
The data simulation and structure reconstruction were performed on an Nvidia Tesla P100 GPU, which has 16 GB RAM. The computation of the micrograph autocorrelations for relevant step sizes took $1.5\times10^2$ seconds on average for a $4096 \times 4096$ micrograph. As for the structure reconstruction, it took a few hours for an instance with a given value of the regularization parameter $\lambda$ to converge. Therefore, if one knows the correct $\lambda$ for some setting, it may be advantageous to use the same $\lambda$ in a similar case. The code is publicly available at https://github.com/tl578/RCT-without-detection.




\referencelist[iucr]

%

\end{document}